\documentclass[conference]{IEEEtran}
\IEEEoverridecommandlockouts
\usepackage{cite}
\usepackage{amsmath,amssymb,amsfonts}
\usepackage{algorithmic}
\usepackage{graphicx}
\usepackage{textcomp}
\usepackage{xcolor}
\usepackage{hyperref}
\usepackage{longtable}
\usepackage{pgfplots}
\usepackage{float}
\usepackage{array}
\pgfplotsset{width=10cm,compat=1.9}
\def\BibTeX{{\rm B\kern-.05em{\sc i\kern-.025em b}\kern-.08em
    T\kern-.1667em\lower.7ex\hbox{E}\kern-.125emX}}

\title{A Personalised Learning Tool for Physics Undergraduate Students Built On a Large Language Model for Symbolic Regression}

\makeatletter
\newcommand{\linebreakand}{%
  \end{@IEEEauthorhalign}
  \hfill\mbox{}\par
  \mbox{}\hfill\begin{@IEEEauthorhalign}
}
\makeatother

\author{\IEEEauthorblockN{1\textsuperscript{st} Yufan Zhu}
\IEEEauthorblockA{\textit{School of Computing} \\
\textit{National University of Singapore}\\
Singapore, Singapore \\
e0773591@u.nus.edu}
\and
\IEEEauthorblockN{2\textsuperscript{nd} Zi-Yu Khoo}
\IEEEauthorblockA{\textit{School of Computing} \\
\textit{National University of Singapore}\\
Singapore, Singapore \\
e0395550@u.nus.edu}
\and
\IEEEauthorblockN{3\textsuperscript{rd} Jonathan Sze Choong Low}
\IEEEauthorblockA{\textit{Agency for Science, Technology and Research (A*STAR)} \\
\textit{Singapore Institute of Manufacturing Technology}\\
Singapore, Singapore \\
sclow@simtech.a-star.edu.sg}
\linebreakand
\IEEEauthorblockN{4\textsuperscript{th} Stéphane Bressan}
\IEEEauthorblockA{\textit{School of Computing} \\
\textit{National University of Singapore}\\
Singapore, Singapore \\
steph@nus.edu.sg}
}

\begin{document}

\maketitle

\begin{abstract}
Interleaved practice enhances the memory and problem-solving ability of students in undergraduate courses. We introduce a personalized learning tool built on a Large Language Model (LLM) that can provide immediate and personalized attention to students as they complete homework containing problems interleaved from undergraduate physics courses. Our tool leverages the dimensional analysis method, enhancing students' qualitative thinking and problem-solving skills for complex phenomena. Our approach combines LLMs for symbolic regression with dimensional analysis via prompt engineering and offers students a unique perspective to comprehend relationships between physics variables. This fosters a broader and more versatile understanding of physics and mathematical principles and complements a conventional undergraduate physics education that relies on interpreting and applying established equations within specific contexts. We test our personalized learning tool on the equations from Feynman's lectures on physics. Our tool can correctly identify relationships between physics variables for most equations,  underscoring its value as a complementary personalized learning tool for undergraduate physics students. 
\end{abstract}


\section{Introduction}

Interleaved practice enhances students' memory and problem-solving ability in undergraduate physics courses~\cite{Samani2021}. This involves students alternating between topics in physics while completing their homework assignments. However, different students learn at different paces, and tutors cannot simultaneously provide complete attention to different students as they work on different topics. The need for a personalized learning tool for each student can be met with Large Language Models (LLMs).

Dimensional analysis, a fundamental concept in physics education, is renowned for its efficacy in cultivating qualitative thinking and unraveling complexities in physics. Its robust application in higher education, as demonstrated in advanced topics like Rayleigh Scattering, underscores its versatility and power. While foundational works by pioneers like Buckingham and Bridgman have cemented its importance, recent studies suggest the potential for its enhanced application in teaching methodologies~\cite{Buckingham1914PhysicallySimilar, Bridgman1922DimensionalAnalysis, Bao2009PhysicsTraining, MarusicErcegSlisko2011, Blasiak2012PhysicsComprehension}. The integration of dimensional analysis into an LLM-driven learning environment further leverages this powerful tool, deepening students' understanding of physics beyond conventional pedagogical boundaries.

LLMs can facilitate the learning of physics from data and the inferencing of laws by leveraging symbolic regression to distill complex physics laws from raw data~\cite{Tenachi2023}. This method aligns well with the historical context of physics discovery, where major theories often emerged from extensive data analysis~\cite{khoo2023, Kepler1609}. Incorporating these algorithms into LLMs allows for a more data-driven and exploratory approach to learning physics, enabling students to infer and understand the underlying principles from empirical data.

In this context, our research introduces a personalized learning tool built on an LLM that leverages dimensional analysis and symbolic regression. This tool is designed to cater to the individual learning pace of students, guiding them through interleaved practice in undergraduate physics courses. Our approach aims to enhance the conventional physics education model, which often limits students to memorizing and applying established equations within specific contexts.

We experimentally evaluate our personalized learning tool by applying it to equations drawn from Feynman's lectures on physics \cite{Feynman1963Mechanics, Feynman1963Quantum, Feynman1963Electromagnetism}. Our tool possesses a remarkable capacity to discover and establish relationships between various physics variables accurately. This efficacy was observed across the majority of the equations tested, which underscores the tool's potential as a beneficial resource in undergraduate physics education. Our tool offers a customized learning experience, adapting to individual student needs and learning styles. Consequently, this tool stands as a valuable complement to traditional teaching methods, enhancing students' grasp of complex physics concepts and their application in various scenarios.

\section{Background}

\subsection{Symbolic Regression}
Symbolic regression is a process that seeks to discover an explicit symbolic formula, representing a mapping from a given set of numerical input-output pairs \(\{x_i, y_i\}_{i=1}^M\), where each \(x_i\) is a vector in a \(d\)-dimensional space (\(x_i \in \mathbb{R}^d\)) and each \(y_i\) is a corresponding output in a \(q\)-dimensional space (\(y_i \in \mathbb{R}^q\)). The core objective is to find a symbolic expression, denoted as \(f\), that accurately describes the mathematical relationship \(y = f(x)\) inherent in the data~\cite{petersen2021deep}. Unlike traditional regression models that might yield complex numerical algorithms, symbolic regression strives to provide a clear, human-readable formula that captures the intricacies and underlying patterns of the dataset.

Symbolic regression is an NP-hard problem, requiring a complex search through an extensive space of mathematical expressions to find the one that best fits a dataset. This complexity involves not just identifying the appropriate mathematical operations but also their optimal arrangement and combination. The resultant expression, \(f\), provides a model that reveals insights into the underlying relationship between variables \cite{Valcarcel2022SymbolicRegressionNP}.

\subsection{Dimensional Analysis}
Dimensional analysis is a powerful tool in physics. It is used to simplify complex problems by focusing on the dimensions of physical quantities. The dimensions of a physical quantity \(Q\) are expressed as \(\left[Q\right] = M^a L^b T^c\), where \(\left[Q\right]\) denotes the dimensions of \(Q\) in terms of mass (M), length (L), and time (T). The utility of dimensional analysis extends beyond mere simplification. Historical contributions like Buckingham's study on physically similar systems and Bridgman's seminal book on dimensional analysis establish the method's foundational role in physics research and education~\cite{Buckingham1914PhysicallySimilar}. 

The importance of dimensional analysis in education is further emphasized by its ability to develop students' intuitive understanding of physical phenomena. 
Furthermore, Blasiak et al. and Taber have also discussed the multi-stage nature of physics comprehension and the importance of sequencing in education, suggesting a need for innovative methods that integrate dimensional analysis more effectively into physics learning~\cite{Blasiak2012PhysicsComprehension, Taber2009TeachingPhysics}.
Dimensional analysis has thus emerged as a crucial tool in developing a deeper and more intuitive understanding of physics, underscoring the need for its effective integration into physics curricula.

Rayleigh's method is an insightful technique within the dimensional analysis literature that seeks to elucidate the relationships between various physical quantities \cite{Rayleigh1915}. It begins by identifying all independent variables that may influence a dependent variable. The method then formulates a functional relationship between these variables, generally as a power product, taking the form:
\begin{equation}
    R = C R_1^a R_2^b R_3^c \ldots R_n^m, \label{rayleigh_method}
\end{equation}
where \( R \) is the dependent variable, \( R_1, R_2, \ldots, R_n \) are the independent variables, \( C \) is a dimensionless constant, and \( a, b, c, \ldots, m \) are the exponents to be determined.

Each variable is expressed in base units, and through the principle of dimensional homogeneity, a set of simultaneous equations is formulated. Solving these equations yields the values of the exponents and forms a dimensionally consistent equation relating the independent and dependent variables. 

\subsection{Large Language Models in Physics Education}
LLMs like the Generative Pre-trained Transformer (GPT) have shown significant potential to revolutionize education, especially in fields like physics. These models can analyze and generate text-based content, aiding in problem-solving and explanation generation~\cite{Liang2023}. In physics education, LLMs have been utilized for solving calculation problems, explaining concepts, and creating new exercises, offering a novel approach to learning and understanding complex topics~\cite{Kieser2023}. However, the effective use of LLMs in physics education requires careful consideration of their capabilities and limitations. Recent studies have highlighted both the potential and the challenges of integrating LLMs into physics education, particularly in terms of accuracy, reliability, and contextual understanding~\cite{Birhane2023}.

\section{Related Works}
\label{sec:lit_review}

\subsection{Symbolic Regression}
\label{subsec:symbolic_regression}

Symbolic regression aims to find mathematical models fitting empirical data, evolving significantly since its early days \cite{gerwin1974information}. Current algorithms are mainly heuristic or exact, with heuristic approaches prioritizing speed and exact methods focusing on accuracy \cite{virgolin2022symbolic}.

Conventional symbolic regression uses genetic programming, an evolutionary technique derived from genetic algorithms and inspired by natural selection \cite{willis1997genetic}. Pioneered by Holland \cite{holland1975adaptation}, and extended by Cramer \cite{cramer1985representation}, this method involves evolving populations of candidate programs, modified through genetic operations like crossover and mutation. Koza's work further demonstrated the effectiveness of genetic algorithms in diverse problem domains, including symbolic regression \cite{koza1994genetic}


Neural networks have revolutionized symbolic regression, leveraging their strength in gradient-based optimization and handling high-dimensional data \cite{biggio2021neural,kubalik2023physically} to enable the discovery of intricate parametric equations\cite{chen2020learning,zhang2022deep}. 

    


Physics-informed symbolic regression integrates physical principles to enhance expression discovery. A key development in this domain is AI Feynman, which employs neural networks to streamline equation discovery, effectively re-discovering equations from the Feynman Lectures on Physics and serving as a benchmark in the field \cite{udrescu2019ai}. Another notable advancement is the Scientist-Machine Equation Detector (SciMED) by Keren et al., an open-source framework that melds scientific knowledge with advanced symbolic regression techniques for identifying physically meaningful symbolic expressions \cite{keren2022computational}.

\subsection{Prompt Engineering for Reasoning}
\label{subsec:prompt_engineering_reasoning}

Reasoning is a structured cognitive process that can be leveraged by LLMs~\cite{galotti1989approaches} to augment their performance in symbolic regression. Effective prompt engineering, which tailors input prompts to guide LLM behavior, is key to maximizing their potential in reasoning tasks \cite{white2023prompt}. This involves single-stage and multi-stage approaches, discussed in Section~\ref{subsubsec:single_stage_approach} and \ref{subsubsec:multi_stage_approach} respectively.

\subsubsection{Single-stage Approach}
\label{subsubsec:single_stage_approach}

The single-stage approach, inspired by the proficiency of LLMs as few-shot reasoners, predominantly employs template-based prompts. Techniques like Contrastive Explanations and POTTER have harnessed prompts to augment commonsense reasoning with pre-trained models \cite{rajagopal2021template}. Notably, the Chain-of-Thought prompting approach introduced by Wei et al. uses a sequence of intermediate reasoning steps within few-shot prompts, making it especially beneficial for multi-step problems \cite{wei2022chain}. Additionally, other techniques have emerged that optimize prompt permutations, examine the impact of exemplar diversity, and incorporate explicit explanations to refine LLM performance in various tasks \cite{lu2021fantastically}. However, the single-stage approach efficacy can be sensitive to the selection of exemplars \cite{webson2021prompt}. 


\subsubsection{Multi-stage Approach}
\label{subsubsec:multi_stage_approach}

The multi-stage approach in prompt engineering, unlike its single-stage counterpart, employs a series of input-output exchanges to better guide LLMs in complex reasoning tasks. This method was substantiated by the foundational works of Kazemi et al. and Creswell and Shanahan, who conceptualized reasoning into sub-modules and developed a selection inference framework, respectively \cite{kazemi2022lambada, creswell2022faithful}. Press et al. advanced this notion by integrating follow-up questions and intermediate answers, thus bridging the compositionality gap in LLMs \cite{press2022measuring}

In recent years, the multi-stage approaches have markedly shaped research, giving rise to various state-of-the-art techniques. This includes Generated-knowledge  Prompting, which enhances LLMs by incorporating external knowledge for improved commonsense reasoning \cite{liu2021generated}. Alongside, Least-to-Most Prompting sequentially dissects complex problems, aiding in tasks that require progressive reasoning \cite{zhou2022least}. 
Adding to the repertoire, Self-Refine Prompting iteratively refines outputs through feedback loops, thus enhancing performance across various tasks \cite{madaan2023selfrefine}.
Each of these techniques represents a significant leap in the ongoing evolution of multi-stage approaches in LLMs.

\section{Methodology}
\label{sec:methodology}

Our methodology uses custom prompts to guide an LLM in generating dimensionally consistent physics equations. We empirically compare different methodologies that use single-stage or multiple-stage prompting approaches and evaluate their efficacy in guiding the model's output. Additionally, we have integrated established theorems, such as Rayleigh's method, to enrich the model's ability to generate physically meaningful expressions, thereby enhancing its utility in both educational and research contexts.

\subsection{Single-stage General Dimensional Analysis Method}
\label{subsec:single_genenral_dimensional_analysis_method}

This method utilizes a single-stage prompting approach for symbolic regression with an LLM. The LLM is presented with both regressors and response variables along with their respective dimensions, and a dataset containing variable values. In this single stage, the LLM is instructed to simultaneously propose dimensionally consistent equations through dimensional analysis and determine the coefficients of the proposed equation by fitting the data.

\begin{quote}
\textbf{Sample Prompt} \\
\textbf{Regressors:} \\
\( X1: [I][T] \) \\
\( X2: [M][L][T]^-3[I]^-1 \) \\
\textbf{Response Variable:} \\
\( Y: [M][L][T]^-2 \) \\
Propose a dimensionally consistent equation form for the response variable as a function of the given regressors with unknown coefficients.
\begin{table}[h]
    \centering
    \begin{tabular}{|ccc|}
        \hline
        X1 & X2 & Y \\
        \hline
        4.230094105945206 & 4.989988020338535 & 21.10811891357122 \\
        1.7835543769151454 & 4.879775762873631 & 8.703345420237708 \\
        & ... & \\
        \hline
    \end{tabular}
\end{table}
\newline
Based on the data set given, derive the unknown coefficients for the equation form proposed in the previous step.
\end{quote}

\subsection{Two-stage General Dimensional Analysis Method}
\label{subsec:two_genenral_dimensional_analysis_method}

This method employs a two-stage approach to prompt symbolic regression. In Stage 1, the LLM is presented with both regressors and response variables, including their respective dimensions. The primary objective is to generate dimensionally consistent equations through dimensional analysis. Following the generation of the equation form, Stage 2 involves supplying LLMs with a dataset containing variable values to determine coefficients.

\begin{quote}
\textbf{Sample Prompt for Stage One} \\
\textbf{Regressors:} \\
\( X1: [I][T] \) \\
\( X2: [M][L][T]^-3[I]^-1 \) \\
\textbf{Response Variable:} \\
\( Y: [M][L][T]^-2 \) \\
Propose a dimensionally consistent equation form for the response variable as a function of the given regressors with unknown coefficients.

\textbf{Sample Prompt for Stage Two} \\
\textbf{Data Points Table:} \\
\begin{table}[h]
    \centering
    \begin{tabular}{|ccc|}
        \hline
        X1 & X2 & Y \\
        \hline
        4.230094105945206 & 4.989988020338535 & 21.10811891357122 \\
        1.7835543769151454 & 4.879775762873631 & 8.703345420237708 \\
        & ... & \\
        \hline
    \end{tabular}
\end{table}
\newline
Based on the data set given, derive the unknown coefficients for the equation form proposed in the previous step.
\end{quote}

\subsection{Single-stage Rayleigh's Method Analysis Method}
\label{subsec:single_rayleigh_method_method}

This method utilizes a single-stage prompting approach, with the integration of Rayleigh's Method for Dimensional Analysis \ref{rayleigh_method}. The LLM is presented with both regressors and response variables along with their respective dimensions, and a dataset containing variable values. The LLM is instructed to simultaneously propose dimensionally consistent equations through Rayleigh's Method and determine the coefficients of the proposed equation by fitting the data.

\begin{quote}
\textbf{Sample Prompt} \\
\textbf{Regressors:} \\
\( X1: [I][T] \) \\
\( X2: [M][L][T]^-3[I]^-1 \) \\
\textbf{Response Variable:} \\
\( Y: [M][L][T]^-2 \) \\
Propose an equation \(Y = C \cdot X_1^{a_1} \cdot X_2^{a_2} \cdot \ldots \cdot X_n^{a_n}\) using Rayleigh's Method, ensuring dimensional consistency. Solve for the unknown exponents \(a_1, a_2, \ldots, a_n\) to make the equation dimensionally homogeneous.

\textbf{Data Points Table:} \\
\begin{table}[h]
    \centering
    \begin{tabular}{|ccc|}
        \hline
        X1 & X2 & Y \\
        \hline
        4.230094105945206 & 4.989988020338535 & 21.10811891357122 \\
        1.7835543769151454 & 4.879775762873631 & 8.703345420237708 \\
        & ... & \\
        \hline
    \end{tabular}
\end{table}
\newline
Based on the data set given, derive the unknown coefficients for the equation form proposed in the previous step.
\end{quote}

\subsection{Two-stage Rayleigh's Method Analysis Method}
\label{subsec:two_rayleigh_method_method}

This method employs a two-stage approach to prompt symbolic regression. In Stage 1, the LLM is presented with both regressors and response variables, including their respective dimensions. The primary objective is to generate dimensionally consistent equations through Rayleigh's Method. Following the generation of the equation form, Stage 2 involves supplying LLMs with a dataset containing variable values to determine coefficients.

\begin{quote}
\textbf{Sample Prompt for Stage One} \\
\textbf{Regressors:} \\
\( X1: [I][T] \) \\
\( X2: [M][L][T]^-3[I]^-1 \) \\
\textbf{Response Variable:} \\
\( Y: [M][L][T]^-2 \) \\
Propose an equation \(Y = C \cdot X_1^{a_1} \cdot X_2^{a_2} \cdot \ldots \cdot X_n^{a_n}\) using Rayleigh's Method, ensuring dimensional consistency. Solve for the unknown exponents \(a_1, a_2, \ldots, a_n\) to make the equation dimensionally homogeneous.

\textbf{Sample Prompt for Stage Two} \\
\textbf{Data Points Table:} \\
\begin{table}[h]
    \centering
    \begin{tabular}{|ccc|}
        \hline
        X1 & X2 & Y \\
        \hline
        4.230094105945206 & 4.989988020338535 & 21.10811891357122 \\
        1.7835543769151454 & 4.879775762873631 & 8.703345420237708 \\
        & ... & \\
        \hline
    \end{tabular}
\end{table}
\newline
Based on the data set given, derive the unknown coefficients for the equation form proposed in the previous step.
\end{quote}

\section{Performance Evaluation}
\subsection{Experiment Setup and Evaluation Metrics}

In our study, we utilized the GPT-4 model, serving as the LLM, and integrated it with a Wolfram Alpha Plugin. This combination significantly bolstered the model's computational abilities, particularly in performing mathematical and symbolic operations, essential for the scope of our research. 
To prevent the model from relying on prior physics knowledge, variables are obfuscated with generic names, e.g., \(E = q_2 \times E_f\) becomes \(Y = X_1 \times X_2\). 
 
Our study utilizes 26 equations from the Feynman Lectures on Physics for undergraduate physics students~\cite{udrescu2019ai}.
We focus on the equations delineated by Udrescu et al. that are solvable by dimensional analysis~\cite{udrescu2019ai}. 
For all regression tasks, we used 20 data points per equation.

We specify the methodologies and metrics employed to assess the effectiveness and efficiency of the LLM. 

The first metric comprises two counts which assesses the LLM's ability to suggest and regress equation forms efficiently and accurately. If the LLM fails to suggest the expected form, the same prompt is iteratively used until success. Therefore a smaller count is better. The first count is the number of prompts required for the LLM to suggest a dimensionally consistent equation. The second is the number of prompts needed to accurately regress the coefficients.

The second metric is the Mean Absolute Percentage Error (MAPE). It is an assessment of fit. The LLM regresses the coefficients, regardless of the dimensional consistency of the initially proposed model. The MAPE is averaged over all 26 equations in the study. 

\subsection{Performance Evaluation}

\begin{table*}[htbp]
\caption{Comparative Analysis of Methodologies}
\begin{center}
\begin{tabular}{|m{1.5cm}|m{2.5cm}|m{1.7cm}|m{1.7cm}|m{2.5cm}|m{1.7cm}|m{1.7cm}|m{1.5cm}|}
\hline
 & \multicolumn{7}{c|}{\textbf{Methodology}} \\
\cline{2-8} 
 & \centering\textbf{Single-stage General Dimensional Analysis} & \multicolumn{2}{m{3.4cm}|}{\centering\textbf{Two-stage General Dimensional Analysis}} & \centering\textbf{Single-stage Rayleigh's Method Analysis} & \multicolumn{2}{m{3.4cm}|}{\centering\textbf{Two-stage Rayleigh's Method Analysis}} & \centering\textbf{AI Feynman} \tabularnewline
\hline
\centering\textbf{Successfully discovered in One Shot} & \centering{17/26} & \multicolumn{2}{|c|}{22/26} &  \centering{17/26} & \multicolumn{2}{|c|}{\centering 23/26} & \multicolumn{1}{|c|}{\centering{26/26}} \\
\hline
\centering\textbf{Average No. of Prompts Needed to} & \centering Regress expected equation & \centering Propose dimensionally consistent equation & \centering Regress expected coefficients & \centering Regress expected equation & \centering Propose dimensionally consistent equations & \centering Regress expected coefficients & \centering N.A. \tabularnewline
\cline{2-8} 
 & \centering1.885 &  \centering1.154 & \centering1.192 & \centering1.308 & \centering1 & \centering1.192 & \centering N.A.\tabularnewline
\hline
\centering\textbf{MAPE} & \centering105.86\% &  \multicolumn{2}{c|}{\centering22.88\%} & \centering9.95\% & \multicolumn{2}{c|}{\centering3.77\%} & \centering $\approx 0 \%$ \tabularnewline
\hline
\centering\textbf{Average Time Taken } & \centering 2m33s & \centering 56s & \centering 2m42s & \centering 2m17s & \centering 1m6s& \centering 2m40s & \centering 18m50s \tabularnewline
\hline
\end{tabular}
\label{tab:method_comparison}
\end{center}
\end{table*}

We assess the four proposed methods utilizing the two aforementioned metrics. The findings, detailed in Table \ref{tab:method_comparison}, underscore the efficiency of LLMs in single-stage prompt resolutions. The success rate for the Single-stage General Dimensional Analysis Method underscores the robust capability of LLM in symbolic regression tasks even within the confines of a single-stage process. Furthermore, the Single-stage Rayleigh's Method Analysis Method requires fewer prompts on average than the Single-stage General Dimensional Analysis Method. The incorporation of an established dimensional analysis approach enhances the LLM's performance further. Lastly, the Two-Stage Rayleigh's Method Analysis Method had the highest success rate. The combination of an established dimensional analysis approach and a multi-stage prompting technique yields the best results. This refinement underscores the effectiveness of our methodology, particularly in tackling sophisticated symbolic regression tasks, demonstrating its practical utility in complex computational scenarios.

In comparison to the baseline symbolic regression AI Feynman, Table \ref{tab:method_comparison}shows that AI Feynman has a higher success rate but is very time-intensive. In contrast, our four methodologies consistently discovered equations within 5 minutes. This efficiency makes our tool more user-friendly and accessible, as students can receive prompt attention and guidance on physics problems, to match their learning paces.

\section{Conclusion}
Our investigation into the application of LLMs for symbolic regression with dimensional analysis tasks in physics has yielded promising results. We observed that both the General Dimensional Analysis Approach and Rayleigh's Method Analysis Approach have proven effective, delivering equations with good accuracy. Notably, our study reveals that two-staged prompting approaches outperform single-stage prompting, underscoring the importance of structured problem-solving and systematic analysis in leveraging LLMs for educational purposes.
Furthermore, the integration of established dimensional analysis methodology such as Rayleigh's Method into prompts has been shown to further enhance the effectiveness of symbolic regression. This synergy allows for a more refined exploration of complex physical phenomena, providing deeper insights that are invaluable in an educational context. 

In conclusion, our research highlights the potential role of LLMs as innovative and accessible educational tools in undergraduate physics education. By harnessing the power of prompt engineering and combining it with scientific knowledge like Rayleigh's Method, LLMs facilitate personalized and data-driven learning experiences, allowing for a nuanced understanding of physical phenomena. Our approach not only deepens students' understanding of physics concepts but also nurtures essential critical thinking and analytical skills, pivotal in today's scientific arena. Consequently, LLMs emerge as valuable assets in the realm of undergraduate physics education, offering novel avenues for learning and exploration. 

\section*{Acknowledgment}
This research is supported by the Singapore Ministry of Education, grant MOE-T2EP50120-0019, and by the National Research Foundation, Prime Minister’s Office, Singapore, under its Campus for Research Excellence and Technological Enterprise (CREATE) program as part of the program Descartes.

\bibliographystyle{IEEEtran}

\end{document}